\documentclass[11pt,twoside]{article}
\usepackage[centertags]{amsmath}
\usepackage{amsfonts}
\usepackage{amssymb}
\usepackage{amsthm}
\usepackage{newlfont}
\usepackage{color}

\date{\color{green}  2013 November}
\def \n {\noindent}
\usepackage{fancyhdr}
\newcommand{\hf}{\hfill $\diamondsuit$}
\begin{document}
\centerline{\bf \bf{\color{green} :::::::}}

\begin{center}
{\bf{\color{red}{\it June  2014}}}\\

***\\

{\Large{\color{blue}{\bf On the chaoticity of some tensor product weighted backward shift operators acting on some tensor product Fock-Bargmann spaces}}}\\
\end{center}

\begin{center}
{\bf Abdelkader INTISSAR  $^{(*)}$ $^{(**)}$}\\
\end{center}

\begin{center}
\scriptsize{(*)Equiped'Analyse spectrale, UMR-CNRS n: 6134, Universit\'e de
Corse, Quartier Grossetti, 20 250 Cort\'e-France \\
 T\'el: 00 33 (0) 4 95 45 00 33\\
 Fax: 00 33 (0) 4 95 45 00 33\\
 e.mail:intissar@univ-corse.fr\\
\parskip = 3pt
(**)Le Prador,129 rue du commandant Rolland, 13008 Marseille-France}
\end{center}

\quad\\

\begin{center}

\fbox{\rule[-0.4cm]{0cm}{1cm} \n {\color{red}{\bf Abstract}}}\\

\end{center}

\n {\it \small{ In Advances in Mathematical Physics (2011) we showed that the weighted shift $z^{p}\frac{d^{p+1}}{dz^{p+1}}$ $(p=0, 1, 2, .....)$ acting on classical Bargmann space $\mathbb{B}_{p}$ is chaotic operator.\\

\n In Journal of Mathematical physics (2014), we constructed an chaotic weighted shift $\mathbb{M}^{*^{p}}\mathbb{M}^{p+1}$ $(p=0, 1, 2, .....)$ on some lattice Fock-Bargmann $\mathbb{E}_{p}^{\alpha}$ generated by the orthonormal basis $\displaystyle{e_{m}^{(\alpha,p)}(z) = e_{m}^{\alpha} ; m=p, p+1, .....}$ where \\

\n $\displaystyle{e_{m}^{\alpha}(z) = (\frac{2\nu}{\pi})^{1/4}e^{\frac{\nu}{2}z^{2}}e^{-\frac{\pi^{2}}{\nu}(m +\alpha)^{2} +2i\pi(m +\alpha)z}; m \in \mathbb{N}}$ with $\nu, \alpha$ are real numbers; $\nu > 0$,\\

\n $\mathbb{M}$ is an weighted shift and $\mathbb{M^{*}}$ is the adjoint of the $\mathbb{M}$.\\

\n In this paper we study the chaoticity of tensor product $\mathbb{M}^{*^{p}}\mathbb{M}^{p+1}\otimes z^{p}\frac{d^{p}}{dz^{p+1}}$ $(p=0, 1, 2, .....)$ \\acting on $\mathbb{E}_{p}^{\alpha}\otimes \mathbb{B}_{p}$.}}{\bf{\color{blue}{\hf}}}\\



\newpage

\begin{center}
\Large{\bf{\color{blue} 1. Introduction and action of $\mathbb{M}^{*^{p}}\mathbb{M}^{p+1}\otimes z^{p}\frac{d^{p+1}}{dz^{p+1}}$ on $\mathbb{E}_{p}^{\alpha}\otimes \mathbb{B}_{p}$}}
\end{center}

 Let $ z = x + iy$; $z \in \mathbb{C}$ and $\nu > 0$, $\alpha$ are fixed real numbers.\\

 We consider the space \\

$ \mathcal{O}_{\alpha}(\mathbb{C}) = \{\phi : \mathbb{C} \rightarrow \mathbb{C}$
entire; $\psi(z+m)= e^{2i\pi\alpha m}e^{\nu(z + \frac{m}{2})m}\psi(z), \forall z \in \mathbb{C}, \forall  m \in \mathbb{N}\}$ \\

 and the Hilbert space \\

 $\mathbb{E}^{\alpha}  = \{ \psi \in \mathcal{O}_{\alpha}(\mathbb{C});  \displaystyle{\int\int_{[0, 1]\times \mathbb{R}}\mid \psi(z)\mid^{2}e^{-\nu\mid z\mid^{2}}dxdy < \infty }\}$ $\hfill { }(1.1)$\\

with inner product,\\

$\displaystyle{< \psi_{1}, \psi_{2} >_{\mathbb{E}^{\alpha}} =  \displaystyle{\int\int_{[0, 1]\times \mathbb{R}}\psi_{1}(z)\overline{\psi}_{2}(z)e^{-\nu\mid z\mid^{2}}dxdy}}$ $\hfill { } (1.2)$\\

and norm,\\

$\displaystyle{\mid\mid \psi \mid\mid_{\mathbb{E}^{\alpha}} = \sqrt{\displaystyle{\int\int_{[0, 1]\times \mathbb{R}}\mid \phi(z)\mid^{2}e^{-\nu\mid z\mid^{2}}dxdy}}}$ $\hfill { } (1.3)$\\

This space is a particular case of $(\Gamma,\chi)$-theta Fock-Bargmann spaces recently constructed by Ghanmi-Intissar in [9]  where it is showed that\\

$\displaystyle{e_{m}^{\alpha}(z) = (\frac{2\nu}{\pi})^{1/4}e^{\frac{\nu}{2}z^{2}}e^{-\frac{\pi^{2}}{\nu}(m +\alpha)^{2} +2i\pi(m +\alpha)z}; m \in \mathbb{N}}$ $\hfill { } (1.4)$\\

is orthonormal basis of $\mathbb{E}^{\alpha}$.{\bf{\color{blue}{\hf}}}\\

{\bf{\color{blue} Remark 1.1}} (fundamental)\\

{\it The explicit construction of orthonormal basis (1.4) of $\mathbb{E}^{\alpha}$ play a fundamental role to know if an operator acting on $\mathbb{E}^{\alpha}$ (in which the polynomials are dense), can be represented as a weighted backward shift.}{\bf{\color{blue}{\hf}}}\\

On $\mathbb{E}^{\alpha}$, we considered in [14] the weight backward shift operator $\mathbb{M}$ defined by\\

$\displaystyle{\mathbb{M}e_{m}^{\alpha} = \gamma_{m-1}e_{m-1}^{\alpha}, m\in \mathbb{N};\quad \mathbb{M}e_{0}^{\alpha} = 0 }$ where $\gamma_{m} = c_{\alpha}e^{\frac{2\pi}{\nu}m}$ and $c_{\alpha}=e^{\frac{\pi}{\nu}+ 2\alpha}$ and we showed  the chaoticity of the operator $ \mathbb{M}^{*^{p}}\mathbb{M}^{p+1}; p=0, 1, 2, ...$ on $\mathbb{E}_{p}^{\alpha}$ where\\

- $\mathbb{E}_{p}^{\alpha} = \{\phi\in \mathbb{E}^{\alpha}; \phi(0) = \phi'(0) = .... \phi^{p-1}(0) = 0$, an orthonormal basis of this space is given by\\

$\displaystyle{e_{m}^{(\alpha,p)}(z) = e_{m}^{\alpha}(z); m = p, p+1, ....}$ $\hfill { }(1.5)$\\

- $\mathbb{M}^{*}$ is adjoint of $\mathbb{M}$ \\

and\\

$\displaystyle{\mathbb{M}^{*^{p}}\mathbb{M}^{p+1}e_{m}^{\alpha,p}(z) = \gamma_{m-1}[\prod_{j=1}^{p}\gamma_{m-1-j}]^{2}e_{m-1}^{\alpha}(z)}$ $\hfill { } (1.6)$\\

 e.g the operator $ \displaystyle{\mathbb{M}^{*^{p}}\mathbb{M}^{p+1}}$ verifies the conditions of the following definition\\

 {\bf{\color{blue}Definition 1.2}}\\

 \quad{\it A linear unbounded densely defined operator $(\mathbb{T}, D(\mathbb{T}))$ on a Banach space $\mathbb{X}$ is called chaotic or Devaney chaotic if the following conditions are met:\\

 1) $\mathbb{T}^{n}$ is closed for all positive integers $n$.\\

 2) there exists an element $\psi \in D(\mathbb{T})^{\infty}$ whose orbit $Orb(\mathbb{T},\psi) = \{\psi, \mathbb{T}\psi, \mathbb{T}^{2}\psi, .....\}$ is dense in $\mathbb{X}$ where $\displaystyle{D(\mathbb{T})^{\infty} = \cap_{n=0}^{\infty}D(\mathbb{T}^{n})}$; such a vector $\psi$ is called a hypercyclic vector for $\mathbb{T}$, the name hypercyclic was motivated by the concept of a cyclic vector from operator theory. In other words, there is no proper closed $\mathbb{T}$-invariant subset of $\mathbb{X}$ containing $\psi$.\\

 3) the set $\{\psi \in \mathbb{X}; \exists \quad m \in \mathbb{N}$ such that $\mathbb{T}^{m}\psi = \psi\}$ of periodic points of operator $\mathbb{T}$ is dense in $\mathbb{X}$.}{\bf{\color{blue}{\hf}}}\\

 In the sequel of this paper, the orbit and the set of periodic points of operator $\mathbb{M}^{*^{p}}\mathbb{M}^{p+1}$ are denoted by $Orb_{\alpha}$ and $\mathbb{U}_{per,\alpha}$ respectively (they are dense in $\mathbb{E}_{p}^{\alpha}$).{\bf{\color{blue}{\hf}}}\\

Now let $\mathbb{B}_{p}$ $(p = 0, 1, ....)$ be the classical Bargmann space defined as a subspace of the space O($\mathbb{C}$) of holomorphic functions on $\mathbb{C}$ such that\\

$ \mathbb{B}_{p} = \{\phi\in O(\mathbb{C}) ; \phi(0) = \phi'(0) = .... \phi^{p-1}(0) = 0$ and $< \phi, \phi >_{\mathbb{B}_{p}} < \infty \}$ $\hfill { } (1.7)$\\

where the pairing $ < ,  >_{\mathbb{B}_{p}} $ is given by\\

$<\phi_{1},\phi_{2}>_{\mathbb{B}_{p}} = \displaystyle {\int_{\mathbb{C}}}\displaystyle{\phi_{1}(z)\overline{\phi_{2}(z)}e^{-\mid z\mid^{2}}dxdy}\hfill { }(1.8)$\\

for all $\phi_{1},\phi_{2} \in O(\mathbb{C})$ and Lebesgue measure $dxdy$ on $\mathbb{C}$.{\bf{\color{blue}{\hf}}}\\

It is easy to verify that the pairing (1.8) defined on the Bargmann space $\mathbb{B}_{p}$ $(p = 0, 1, ....)$ is an inner product and the associated norm is \\

$\mid\mid \phi \mid\mid_{\mathbb{B}_{p}} $= $\sqrt{\displaystyle {\int_{\mathbb{C}}}\displaystyle{\mid\phi(z)\mid^{2}e^{-\mid z\mid^{2}}dxdy}}$ $\hfill { }(1.9)$\\

Now, we can used a theorem of Weierstrass to show that any Cauchy sequence in $\mathbb{B}_{p}$ has a limit $\phi\in O(\mathbb{C})$ and we check that $\phi \in \mathbb{B}_{p}$ and indeed is the limit of the Cauchy sequence in the norm $\mid\mid . \mid\mid_{\mathbb{B}_{p}} $ of $\mathbb{B}_{p}$ induced by the inner product. These steps show that the space $\mathbb{B}_{p}$ is complete and we have\\

i) The classical Bargmann space $\mathbb{B}_{p}$ is a Hilbert space.\\

ii) An orthonormal basis of $\mathbb{B}_{p}$ is given by\\

$\displaystyle{e_{n}^{p}(z) = \frac{z^n}{\sqrt{n!}}; n = p, p+1, ....}$ $\hfill { }(1.10)$\\

On $\mathbb{B}_{p}$ which is the orthogonal of span $\displaystyle{\{e_{n}^{p}; n < p\}}$ in Bargmann space[3]\\

$\displaystyle{\mathbb{B}_{0} = \{\phi : \mathbb{C} \rightarrow \mathbb{C} \quad entire \quad ; \int_{\mathbb{C}}\displaystyle{\mid\phi(z)\mid^{2}e^{-\mid z\mid^{2}}dxdy}\}}$ $\hfill { }(1.11)$\\
with its usual basis:\\

$\displaystyle{e_{n}(z) = \frac{z^n}{\sqrt{n!}}; n = 0, 1, ....}$ $\hfill { }(1.12)$\\

We considered in [15] the annihilator operator $\frac{d}{dz}$ defined by\\

$\displaystyle{\frac{d}{dz}e_{n} = \omega_{n-1}e_{n-1}, n\in \mathbb{N};\quad e_{-1} = 0 }$ where $\omega_{n} = \sqrt{n+1}$ and we showed  the chaoticity of the operator $ z^{p}\frac{d^{p}}{dz^{p+1}}; p=0, 1, 2, ...$ where $z$ is adjoint of $\frac{d}{dz}$ and\\

$\displaystyle{z^{p}\frac{d^{p}}{dz^{p+1}}e_{n}^{p}(z) = \omega_{n-1}[\prod_{j=1}^{p}\omega_{n-1-j}]^{2}e_{n-1}^{p}(z)}$ $\hfill { } (1.13)$\\

 e.g the operator $ \displaystyle{z^{p}\frac{d^{p}}{dz^{p+1}}}$ verifies the conditions of the  definition 1.2 {\bf{\color{blue}{\hf}}}\\

 In the sequel of this paper, the orbit and the set of periodic points of operator $z^{p}\frac{d^{p}}{dz^{p+1}}$ are denoted by $Orb_{p}$ and $\mathbb{U}_{per,p}$ respectively (they are dense in $\mathbb{B}_{p}$).\\

$\mathbb{E}_{p}^{\alpha}$ and $\mathbb{B}_{p}$ are two Hilbert spaces then the tensor product of $\mathbb{E}_{p}^{\alpha}$ and $\mathbb{B}_{p}$ is a new Hilbert space $\mathbb{E}_{p}^{\alpha}\otimes \mathbb{B}_{p}$.(see, e.g.,[28, Theorem 3.12(b)]).{\bf{\color{blue}{\hf}}}\\

The reader is referred to Schatten [27] for the theory of cross-spaces and Kubrisly [20] for a concise introduction to tensor product of bounded operators or to Reed-Simon [23] for tensor products of closed operators on Banach spaces.\\

Below we list a few remarks concerning properties of tensor products, which we will use in the sequel.{\bf{\color{blue}{\hf}}}\\

Define the elementary elements of the space $\mathbb{E}_{p}^{\alpha}\otimes \mathbb{B}_{p}$ as pairs of $\psi \in \mathbb{E}_{p}^{\alpha}$ and $\phi \in \mathbb{B}_{p}$ and written as $\psi\otimes\phi$ where\\

$\psi\otimes\phi$:$\displaystyle{\mathbb{E}_{p}^{\alpha}\times \mathbb{B}_{p} \rightarrow \mathbb{C}}$\\

\quad\quad $\displaystyle{(f, g) \rightarrow \psi\otimes\phi(u,v) = < \psi, u >_{\mathbb{E}_{p}^{\alpha}}<\phi, v >_{\mathbb{B}_{p}}}$ $\hfill { } (1.14)$\\

$\psi\otimes\phi$ is called single tensor product and we observe that the single $0\otimes0$ coincides with $\psi\otimes0$ and $0\otimes\phi$ and the natural map $(\psi, \phi) \rightarrow \psi\otimes\phi$ is not injective.\\

For $\lambda \in \mathbb{C}$, one identifies $\lambda(\psi\otimes \phi) = (\lambda\psi)\otimes \phi =\psi\otimes (\lambda\phi)$, and considers formal sums of vectors of these elementary vectors. One takes the inner product of two elementary vectors in $\mathbb{E}_{p}^{\alpha}\otimes \mathbb{B}_{p}$ as the product of the corresponding inner products,\\

$\displaystyle{<\psi_{1}\otimes \phi_{1}, \psi_{2}\otimes \phi_{2} >_{\mathbb{E}_{p}^{\alpha}\otimes \mathbb{B}_{p}} = < \psi_{1}, \psi_{2} >_{\mathbb{E}_{p}^{\alpha}}.<\phi_{1}, \phi_{2} >_{\mathbb{B}_{p}}}$ $\hfill { } (1.15)$\\

One extends this definition by linearity to finite sums of $m$$n$ elementary vectors

$\displaystyle{\Phi = \sum_{i=p}^{m}\sum_{j=p}^{n}a_{ij}\psi_{i}\otimes\phi_{j}}$, where $a_{mn} \in \mathbb{C}$\\

Let $\displaystyle{\Psi = \sum_{k=p}^{m'}\sum_{l=p}^{n'}b_{kl}\psi_{i}\otimes\phi_{j}}$, where $b_{m'n'} \in \mathbb{C}$\\

The inner product of two such vectors $\displaystyle{\Psi}$ and $\displaystyle{\Phi}$ must be linear in $\displaystyle{\Psi}$ and conjugate linear in $\displaystyle{\Phi}$\\

Thus the inner product must have the form:\\

$\displaystyle{<\Psi, \Phi >}_{\mathbb{E}_{p}^{\alpha}\otimes \mathbb{B}_{p}}\displaystyle{ = \sum_{i=p}^{m}\sum_{j=p}^{n} \sum_{k=p}^{m'}\sum_{l=p}^{n'} \overline{c}_{ij}c_{kl} <\psi_{k}, \psi_{l} >_{\mathbb{E}_{p}^{\alpha}}.<\phi_{m}, \phi_{n} >_{\mathbb{B}_{p}}}$ $\hfill { } (1.16)$\\

The condition that this form make $\mathbb{E}_{p}^{\alpha}\otimes \mathbb{B}_{p}$ into
a pre-Hilbert space is the statement that $ \displaystyle{0 \leq  < \Psi, \Psi >_{\mathbb{E}_{p}^{\alpha}\otimes \mathbb{B}_{p}}}$, with vanishing only possible if $ \displaystyle{\Psi = 0}$. In other words, the form (1.8) is positive definite on $\displaystyle{(\mathbb{E}_{p}^{\alpha}\otimes \mathbb{B}_{p})\times (\mathbb{E}_{p}^{\alpha}\otimes \mathbb{B}_{p})}$. In this case, the algebraic tensor product $\displaystyle{\mathbb{E}_{p}^{\alpha}\otimes \mathbb{B}_{p}}$ is a pre-Hilbert space that can be completed to a Hilbert space that we call $\displaystyle{\mathbb{E}_{p}^{\alpha}\otimes \mathbb{B}_{p}}$.{\bf{\color{blue}{\hf}}}\\

 In this work as $\displaystyle{e_{m}^{\alpha} \in \mathbb{E}_{p}^{\alpha}}$ is an orthonormal basis of $\mathbb{E}_{p}^{\alpha}$ and $\displaystyle{e_{n}^{p} \in \mathbb{B}_{p}}$ is an orthonormal base of $\mathbb{B}_{p}$ then $\displaystyle{e_{m}^{\alpha}\otimes e_{n}^{p}}$ is an orthonormal basis for $\mathbb{E}_{p}^{\alpha}\otimes \mathbb{B}_{p}$.\\

 Now let be tow linear operators $T_{1}$ with domain $D(T_{1})$ on $\mathbb{E}_{p}^{\alpha}$ and $T_{2}$ with domain $D(T_{2})$ on $\mathbb{B}_{p}$ respectively, we define the tensor product operator $T_{1}\otimes T_{2}$ of $T_{1}$ and $T_{2}$ on $\mathbb{E}_{p}^{\alpha}\otimes \mathbb{B}_{p}$ by:\\

$\displaystyle{(T_{1}\otimes T_{2}) (\psi\otimes\phi) = (T_{1}\psi)\otimes(T_{1}\phi)}$, $\hfill { } (1.17)$\\

$\psi \in D(T_{1})$ and $\phi \in D(T_{2})$ \\

and extends this definition by linearity to all of $D(T_{1})\otimes D(T_{1})$. As a consequence by adapting Theorem 7.18 in [8] to unbounded operators we get,\\

i) $\displaystyle{(T_{1}\otimes T_{2})(T_{1}^{'}\otimes T_{2}^{'}) = (T_{1}T_{1}^{'})\otimes (T_{2}T_{2}^{'})}$ $\hfill { } (1.18)$\\

on $D(T_{1}T_{1}^{'})\otimes D(T_{1}T_{1}^{'})$ \\

ii) $\displaystyle{(T_{1}\otimes T_{2})^{*} = (T_{1}^{*}\otimes T_{2}^{*})}$ $\hfill { } (1.19)$\\

on $\displaystyle{D((T_{1}\otimes T_{2})^{*}) \cap D(T_{1}^{*})\otimes D(T_{2}^{*})}$ \\

iii) $\displaystyle{(T_{1}\otimes T_{2})^{*}( T_{1} \otimes T_{2}) = (T_{1}^{*}T_{1})\otimes (T_{2}^{*}T_{2})}$ $\hfill { } (1.20)$\\

on $\displaystyle{D((T_{1}\otimes T_{2})^{*}(T_{1} \otimes T_{2})) \cap D(T_{1}^{*}T_{1})\otimes D(T_{2}^{*}T_{2})}$. \\

The matrix elements of $\displaystyle{T_{1}\otimes T_{2}}$ in the basis $\displaystyle{\{e_{m}^{\alpha,p}\otimes e_{n}^{p}\}}$ can be expressed in terms of the matrix
elements of $\displaystyle{T_{1}}$ in the basis $\displaystyle{\{e_{m}^{\alpha,p}\}}$ and $\displaystyle{T_{2}}$ in the basis $\displaystyle{\{e_{n}^{p}\}}$. {\bf{\color{blue}{\hf}}}\\

In this paper we are concerned with the problem of preserving properties of chaoticity by
tensor product of weighted shift $\mathbb{M}^{*^{p}}\mathbb{M}^{p+1}$ on $\mathbb{E}_{p}^{\alpha}$ with weighted shift $z^{p}\frac{d^{p}}{dz^{p+1}}$ on $\mathbb{B}_{p}$.\\

{\bf{\color{blue} Remark 1.3}}\\

{\it We cannot use the machinery developed by Reed-Simon [23] because the operators $\mathbb{T}_{1}:=\mathbb{M}^{*^{p}}\mathbb{M}^{p+1}$ and $\mathbb{T}_{2}:= z^{p}\frac{d^{p}}{dz^{p+1}}$  have empty resolvent sets on $\mathbb{E}_{p}^{\alpha}$ and $\mathbb{B}_{p}$ because their spectrum $\displaystyle{\sigma(\mathbb{T}_{1}) = \sigma(\mathbb{T}_{2}) = \mathbb{C}}$.\\

Then we show in next section that the operator $\mathbb{T}_{1}\otimes \mathbb{T}_{2}$ verifies the conditions of definition 1.2 of the chaoticity on $\mathbb{E}_{p}^{\alpha}\otimes \mathbb{B}_{p}$.{\bf{\color{blue}{\hf}}}\\

\begin{center}
\Large{\bf{\color{blue} 2. On the chaoticity of $\mathbb{M}^{*^{p}}\mathbb{M}^{p+1}\otimes z^{p}\frac{d^{p+1}}{dz^{p+1}}$ on $\mathbb{E}_{p}^{\alpha}\otimes \mathbb{B}_{p}$}}
\end{center}

In this section, following [14] and [15] we recall that the operators $\mathbb{T}_{1}:=\mathbb{M}^{*^{p}}\mathbb{M}^{p+1}$ and $\mathbb{T}_{2}:= z^{p}\frac{d^{p}}{dz^{p+1}}$ are chaotic on $\mathbb{E}_{p}^{\alpha}$ and on $\mathbb{B}_{p}$ respectively and we are concerned with the problem of preserving  of this property by
tensor product of $\mathbb{T}_{1}$ and $\mathbb{T}_{2}$.{\bf{\color{blue}{\hf}}}\\

The study of the phenomenon of hypercyclicity originates in the papers by Birkoff {\bf [5]} and Maclane {\bf [21]} that show, respectively, that the operators of translation and differentiation, acting on the space of entire functions are hypercyclic. \\

The theories of hypercyclic operators and chaotic operators have been intensively developed for bounded linear operator, we refer to {\bf [2, 7, 10, 11, 12, 13]} and references therein.
{\bf{\color{blue}{\hf}}}\\

{\bf{\color{blue} Remark 2.1}}\\

{\it i) For a bounded operator, Ansari  asserts in  {\bf [1]} that powers of a hypercyclic bounded operator are also hypercyclic.\\

ii) In {\bf [25]} Salas asserts that Weighted backward shifts constitute an important class of operators which is the ''favorite testing ground'' for hypercyclicity and characterizes hypercyclicity of weighted backward shift $\mathbb{T}_{\omega}$ acting on \\

$\displaystyle{l_{p} = \{(x_{n})_{n=0}^{\infty} \in \mathbb{C} ; \sum_{n=0}^{\infty}\mid x_{n}\mid^{p} < \infty\}}$ $( 1\leq  p < +\infty$ or $p=0)$ $\hfill { } (2.1)$\\

where $\mathbb{T}_{\omega}$ is defined by:\\

$\displaystyle{\mathbb{T}_{\omega}(x_{0}, x_{1}, x_{2}, ......) = (\omega_{1}x_{1}, \omega_{2}x_{2}, \omega_{3}x_{3}, ......)}$ $\hfill { } (2.2)$\\

where $(\omega_{1}, \omega_{2}, \omega_{3}, ......)$ is a sequence of numbers.\\

then\\

a) $\mathbb{T}_{\omega}$ is well defined and continuous if and only if $\displaystyle{(\omega_{i})_{i=1}^{\infty} \in l_{\infty}}$\\

b) $\mathbb{T}_{\omega}$ is
hypercyclic on $l_{p}$ if and only if $\displaystyle{Sup_{n\in N}\prod_{i=1}^{\infty}\omega_{i} = \infty}$.}{\bf{\color{blue}{\hf}}}\\

 In [22], Martinez-Gimen$\grave{e}$z and Peris assert, on universality and chaos for tensor products of bounded weighted backward shift operators, the following proposition\\

{\bf {\color{blue} Proposition 2.2}} [22] \\

{\it Let $1 \leq p; q \leq \infty $ and let $\mathbb{T}_{\omega} : l_{p} \rightarrow l_{p}$; $\mathbb{T}_{\varpi} : l_{q} \rightarrow l_{q}$ be two bounded weighted backward shifts. Then
$\displaystyle{\mathbb{T}_{\omega}\otimes\mathbb{T}_{\varpi}: l_{p}\otimes l_{q} \rightarrow l_{p}\otimes l_{q}}$ is hypercyclic on $l_{p}\otimes l_{q}$ if and only if $\displaystyle{Sup_{n\in N}\prod_{i=1}^{\infty}\mid \omega_{i} \varpi_{i}\mid= \infty}$.}{\bf{\color{blue}{\hf}}}\\

{\bf {\color{blue}Remark 2.3}}\\

{\it The tensor product of two hypercyclic operators is not necessarily hypercyclic.}{\bf{\color{blue}{\hf}}}\\

It is sufficient to take the pair of  weights shifts defined by \\

$\displaystyle{\mathbb{T}_{\omega}(x_{0}, x_{1}, x_{2}, ......) = (2x_{1}, \frac{1}{2}x_{2}, \frac{1}{2}x_{3}, 2x_{4},2x_{5}......)}$ $\hfill { } (2.3)$\\

and\\

$\displaystyle{\mathbb{T}_{\varpi}(x_{0}, x_{1}, x_{2}, ......) = (\frac{1}{2}x_{1}, 2x_{2}, 2x_{3}, \frac{1}{2}x_{4}, \frac{1}{2}x_{5}......)}$ $\hfill { } (2.4)$\\

so by using the characterization of Salas [25] we deduce that $\mathbb{T}_{\omega}$ and $\mathbb{T}_{\varpi}$ are hypercyclic. Clearly $\displaystyle{Sup_{n\in N}\prod_{i=1}^{\infty}\mid \omega_{i} \varpi_{i}\mid= 1}$ for all $n \in \mathbb{N}$ and by the above proposition, we obtain that $\displaystyle{\mathbb{T}_{\omega}\otimes\mathbb{T}_{\varpi}}$ is not hypercyclic.{\bf{\color{blue}{\hf}}}\\

{\bf{\color{blue} Remark 2.4}}\\

{\it i) For an unbounded operator, Salas exhibit in {\bf [24]} an unbounded hypercyclic operator whose square is not hypercyclic.\\

ii) Let $\mathbb{B}$ the classical Bargmann space with orthonormal basis \\

$\{e_{n} = \frac{z^{n}}{\sqrt{n!}}; n =0, 1, . . . $ \\

Define the lowering and raising operators $\mathbb{A}$ and $\mathbb{A}^{*}$ as\\

$\mathbb{A}e_{n} = \sqrt{n}e_{n-1}$ , $\mathbb{A}e_{n} = 0$ (lowering operator) $\hfill { } (2.5)$\\

$\mathbb{A}^{*}e_{n} = \sqrt{n+1}e_{n+1}$ ,  (raising operator) $\hfill { } (2.6)$\\

a) It is well known that the annihilator operator $\mathbb{A}$ acting on classical Bargmann space is chaotic see [15] but $\mathbb{A}^{*}\mathbb{A}$ and $\mathbb{A}^{*}$ + $\mathbb{A}$ are not chaotic where $\mathbb{A}^{*}$ is the creator operator.\\

b) So it is well known that the operators $\mathbb{A}^{*}\mathbb{A}^{2}$ and $\mathbb{A}^{*}\mathbb{A}^{2} + \mathbb{A}^{*^{2}}\mathbb{A}$ acting on classical Bargmann space $B_{p}$ ; $p=0$ are chaotic see [6].\\

Then \\

there exist some operators $\mathbb{T}_{1}$  and $\mathbb{T}_{2}$ acting on Bargmann space such that $\mathbb{T}_{1}$ is chaotic and $\mathbb{T}_{2}$  is not chaotic with $\mathbb{T}_{1} + \mathbb{T}_{2}$ is chaotic, it suffices to take $\mathbb{T}_{1} = \mathbb{A}^{*}\mathbb{A}^{2}$ and
$\mathbb{T}_{2} = \mathbb{A}^{*^{2}}\mathbb{A}$. An complete scattering analysis on $\mathbb{A}^{*}\mathbb{A}^{2} + \mathbb{A}^{*^{2}}\mathbb{A}$ acting on Bargmann space is given in [17].\\

or\\

$\mathbb{T}_{1}$  and $\mathbb{T}_{2}$ acting on Bargmann space such that $\mathbb{T}_{1}$ is chaotic and $\mathbb{T}_{2}$  is not chaotic with $\mathbb{T}_{1} + \mathbb{T}_{2}$ is not chaotic, it suffices to take $\mathbb{T}_{1} = i(\mathbb{A}^{*^{2}}\mathbb{A} + \mathbb{A}^{*}\mathbb{A}^{2})$ with $i^{2} = -1$ and $\mathbb{T}_{2} = \mathbb{A}^{*}\mathbb{A}$. An complete spectral analysis is given in [18] and [19].\\

c) The polynomial operators $P(\mathbb{A}^{*},\mathbb{A})$ acting on classical Bargmann space are an excellent laboratory for the study the phenomenons of hypercyclicity or of chaoticity.\\

d) Generally it is observed that many properties of concept of tensor universality criterion for a sequence of bounded operators are not applicable to a sequence of unbounded operators, in particular to our operators.}\\

The above results show that one must be careful in the formal manipulation of operators with restricted domains. For such operators it is often more convenient to work with vectors rather than with operators themselves.{\bf{\color{blue}{\hf}}}\\

 We will denote by $\displaystyle{D(\mathbb{T}_{1})\otimes D(\mathbb{T}_{2})}$ the set of linear linear combinations of vectors of the $\psi\otimes\phi$ where $\psi \in D(\mathbb{T}_{1})$ and $\phi \in D(\mathbb{T}_{2})$.\\

 As $D(\mathbb{T}_{1})$ and $D(\mathbb{T}_{2})$ are dense in $\mathbb{E}_{p}^{\alpha}$ and in $\mathbb{B}_{p}$ respectively then\\

 $D(\mathbb{T}_{1})\otimes D(\mathbb{T}_{2})$ is dense in $\mathbb{E}_{p}^{\alpha} \times \mathbb{B}_{p}$.\\

We define $\mathbb{T}_{1}\otimes \mathbb{T}_{2}$ by
$(\mathbb{T}_{1}\otimes \mathbb{T}_{2})(\psi \otimes \phi) = \mathbb{T}_{1}\psi \otimes \mathbb{T}_{2}\phi $ $\hfill { } (2.7)$\\

and extend by linearity.{\bf{\color{blue}{\hf}}}\\

{\bf{\color{blue} Lemma 2.5}}\\

{\it
 i) The operator $\mathbb{T}: = \mathbb{T}_{1}\otimes \mathbb{T}_{2} = \mathbb{M}^{*^{p}}\mathbb{M}^{p+1}\otimes z^{p}\frac{d^{p+1}}{dz^{p+1}}$ is closable.\\

 ii) For each positive integer $k$, the operator $\mathbb{T}^{k}$ is a closed.}{\bf{\color{blue}{\hf}}}\\

{\bf Proof}\\

i) Let $\Phi \in D(\mathbb{T}_{1})\otimes D(\mathbb{T}_{2})$ and $Psi$ is any vector in $D(\mathbb{T}_{1}^{*})\otimes D(\mathbb{T}_{2}^{*})$, then\\

 $< \mathbb{T}_{1}\otimes \mathbb{T}_{2}\Phi, \Psi > = < \Phi, \mathbb{T}_{1}^{*}\otimes\mathbb{T}_{2}^{*}\Psi >$ \\

so \\

$D(\mathbb{T}_{1}^{*})\otimes D(\mathbb{T}_{2}^{*}) \subset D((\mathbb{T}_{1}\otimes \mathbb{T}_{2})^{*}))$.\\

As $\mathbb{T}_{1}$ and $\mathbb{T}_{2}$ are closable, $D(\mathbb{T}_{1}^{*})$ and $D(\mathbb{T}_{2}^{*})$ are dense.\\

Therefore, in this case $(\mathbb{T}_{1}\otimes \mathbb{T}_{2})^{*} $ is densely defined which proves that $\mathbb{T}_{1}\otimes \mathbb{T}_{2}$ is closable.\\

ii) As $\mathbb{T}^{k}: =[\mathbb{M}^{*^{p}}\mathbb{M}^{p+1}\otimes z^{p}\frac{d^{p+1}}{dz^{p+1}}]^{k}$ is a closed if and only if its graph $\mathfrak{G}(\mathbb{T}^{k})$ is a closed linear manifold of
$\mathbb{E}_{p}^{\alpha}\otimes \mathbb{B}_{p}$ $\times$ $\mathbb{E}_{p}^{\alpha}\otimes \mathbb{B}_{p}$.\\

Let $(f_{n}, \mathbb{T}^{k}f_{n})$ be an sequence witch converges to some $(f, g)$ in $\mathbb{E}_{p}^{\alpha}\otimes \mathbb{B}_{p}$ $\times$ $\mathbb{E}_{p}^{\alpha}\otimes \mathbb{B}_{p}$.\\

We want to show that $f \in D(\mathbb{T}^{k})$ and $g = \mathbb{T}^{k}f$. To see this, it suffices  to take $f_{n}$ of the form $f_{n} = \psi_{n}\otimes\phi_{n}$ where \\

$\psi_{n}\otimes\phi_{n}$: $\mathbb{E}_{p}^{\alpha}\times \mathbb{B}_{p} \rightarrow \mathbb{C}$\\

\quad \quad $(u, v) \rightarrow \psi_{n}\otimes\phi_{n} (u, v) = < \psi_{n}, u > +  <\phi_{n}, v >$ $\hfill { } (2.8)$\\

Then \\

$\psi_{n}\otimes\phi_{n}$ converges to some $\psi\otimes\phi$ in  $\mathbb{E}_{p}^{\alpha}\otimes \mathbb{B}_{p}$.\\

In particular $\psi_{n}(z)$ converges to  $\psi(z)$ in $\mathbb{C}$ and $\phi_{n}(z')$ converges to  $\phi(z')$ in $\mathbb{C}$.\\

As $\mathbb{T}_{1}^{m}$ and $\mathbb{T}_{2}^{k}$ are closed, then we deduce that\\

 $< \mathbb{T}_{1}^{k}\psi_{n}, u > \rightarrow  <\mathbb{T}_{1}^{k}\psi, u > \quad \forall\quad u \in  \mathbb{E}_{p}^{\alpha}$ $\hfill { } (2.9)$\\

 and\\

 $< \mathbb{T}_{2}^{k}\psi_{n}, v > \rightarrow  < \mathbb{T}_{2}^{k}\psi, v > \quad \forall\quad v\in  \mathbb{B}_{p}$ $\hfill { } (2.10)$\\

and\\

$< \mathbb{T}_{1}^{k}\psi_{n}, u > < \mathbb{T}_{2}^{k}\psi_{n}, v > \rightarrow <\mathbb{T}_{1}^{k}\psi, u >< \mathbb{T}_{2}^{k}\psi, v > \forall u \in  \mathbb{E}_{p}^{\alpha}, \forall v\in  \mathbb{B}_{p}$ $\hfill { } (2.11)$\\

Now as $\mathbb{T}^{k} = \mathbb{T}_{1}^{k}\otimes\mathbb{T}_{2}^{k}$ we deduce from (2.11) that\\

$\mathbb{T}^{k}(\psi_{n}\otimes\phi_{n})$ converges to $\mathbb{T}^{k}(\psi\otimes\phi)$ then $\psi\otimes\phi \in D(\mathbb{T}^{k})$ and $g = \mathbb{T}^{k}(\psi\otimes\phi)$.{\bf{\color{blue}{\hf}}}\\

{\bf{\color{blue} Remark 2.6}}\\

{\it a) In the proof of i), we observe that if two unbounded operators are closable then their tensor product is closable also. e.g the property to be closable is preserved by tensor product.\\

b) We can exhibit a closed operator whose square is not. For example, the operator acting on $L_{2}[0, 1] \times L_{2}[0, 1]$ defined by \\

$\mathbb{T}(u, v)(x) = (v'(x), f(x)v(0))$ with domain $D(\mathbb{T}) =L_{2}[0, 1]\times H_{1}[0, 1]$ $\hfill { } (2.12)$\\

 where $v'(x)$ is the derivative of $v(x)$ and $f$ is a function in $H_{1}[0, 1]$ with $f(0) = 1$, $H_{1}[0, 1]$ is the classical Sobolev space.\\

 Then $\mathbb{T}$, is a closed operator and $D(\mathbb{T}^{2}) = D(\mathbb{T})$, where $D(\mathbb{T}^{2})$ is the domain of $\mathbb{T}^{2}$ but the operator $\mathbb{T}^{2}$ is not closed and has not closed extension.This operator can, for example, justify the first assumption  of the Definition 1.2 for the unbounded linear operators.}{\bf{\color{blue}{\hf}}}\\

Sufficient conditions for the hypercyclicity of an unbounded operator are given in the following B$\grave{e}$s-Chan-Seubert theorem:\\

{\bf{\color{blue} Theorem 2.7}}\quad (B$\grave{e}$-Chan-Seubert [4])\\

{\it Let $\mathbb{X}$ be a separable infinite dimensional Banach, and let $\mathbb{T}$ be a densely defined linear operator on $\mathbb{X}$. Then, $\mathbb{T}$ is hypercyclic if \\

i) $\mathbb{T}^{k}$ is a closed operator for all positive integers $k$\\

ii) there exists a dense subset $\mathbb{F}$ of the domain $D(\mathbb{T})$ of $\mathbb{T}$ and a (possibly nonlinear and discontinuous) mapping $\mathbb{S} : \mathbb{F} \rightarrow  \mathbb{F}$ so that $\mathbb{T}\mathbb{S}$ is the identity on $\mathbb{F}$ and $\mathbb{T}^{k}, \mathbb{S}^{k} \rightarrow  0 $ pointwise on $\mathbb{F}$ as $k \rightarrow  +\infty$.}{\bf{\color{blue}{\hf}}}\\

In [14] and [15] we showed that $\mathbb{T}_{1}:=\mathbb{M}^{*^{p}}\mathbb{M}^{p+1}$ and $\mathbb{T}_{2}:= z^{p}\frac{d^{p}}{dz^{p+1}}$ are chaotic on $\mathbb{E}_{p}^{\alpha}$ and on $\mathbb{B}_{p}$ in particular they are hypercyclic.

 We verify now that the operator $\mathbb{T}= \mathbb{T}_{1}\otimes\mathbb{T}_{2}= \mathbb{M}^{*^{p}}\mathbb{M}^{p+1}\otimes z^{p}\frac{d^{p+1}}{dz^{p+1}}$ on $\mathbb{E}_{p}^{\alpha}\otimes \mathbb{B}_{p}$ satisfies the hypercyclicity criterion, as quoted above. {\bf{\color{blue}{\hf}}}\\

{\bf{\color{blue} Lemma 2.8}}\\

{\it Let $\mathbb{T}_{1}= \mathbb{M}^{*^{p}}\mathbb{M}^{p+1}$ with domain $D(\mathbb{T}_{1}) = \{\psi \in \mathbb{E}_{p}^{\alpha}; \mathbb{T}_{1}\psi \in \mathbb{E}_{p}^{\alpha}\}$  where\\

 $\mathbb{T}_{1}e_{m}^{\alpha,p} =\gamma_{m}^{\alpha,p}e_{m-1}^{\alpha,p}$ with $\gamma_{m}^{\alpha,p} = \gamma_{m-1}[\prod_{j=1}^{p}\gamma_{m-1-j}]^{2}$; $\gamma_{m} = c_{\alpha}e^{\frac{2\pi}{\nu}m}$ and\\

 $c_{\alpha}=e^{\frac{\pi}{\nu}+ 2\alpha}$ for $m \geq p \geq 0$\\

and\\

 $\mathbb{T}_{2}= z^{p}\frac{d^{p+1}}{dz^{p+1}}$ with domain $D(\mathbb{T}_{2}) = \{\phi \in \mathbb{B}_{p}; \mathbb{T}_{2}\phi \in \mathbb{B}_{p}\}$  where\\

 $\mathbb{T}_{2}e_{n}^{p} =\omega_{n}{p}e_{n-1}^{p}$ with $\omega_{n}^{p} = \sqrt{n+1}\frac{n!}{(n-p)!}$ for $n \geq p \geq 0$\\

 Then \\

$\mathbb{T} = \mathbb{T}_{1}\otimes\mathbb{T}_{2}$ with domain $D(\mathbb{T}) = D(\mathbb{T}_{1})\otimes D(\mathbb{T}_{2})$ is hypercyclic.}{\bf{\color{blue}{\hf}}}\\

{\bf Proof}\\

Let $\displaystyle{\mathbb{F}_{\alpha} = \{\psi_{k} = \sum_{m=p}^{k}a_{m}e_{m}^{\alpha,p}\}}$ and $\displaystyle{\mathbb{F} = \{\psi_{k} = \sum_{n=p}^{k}b_{n}e_{n}^{p}\}}$ these spaces are dense in\\

$\mathbb{E}_{p}^{\alpha}$ and $\mathbb{B}_{p}$ respectively.\\

Let $\mathbb{S}_{1,p}: \mathbb{F}_{\alpha} \rightarrow \mathbb{F}_{\alpha}$
 defined by $\displaystyle{\mathbb{S}_{1,p}e_{m,p}^{\alpha} = \frac{1}{\gamma_{m}^{\alpha,p}}e_{m+1,p}^{\alpha}; m \geq p \geq 0}$\\

 and\\

 $\mathbb{S}_{2,p}: \mathbb{F} \rightarrow \mathbb{F}$ defined by $\displaystyle{\mathbb{S}_{1,p}e_{m}^{\alpha} = \frac{1}{\omega_{n}^{p}}e_{n+1}^{p}; n \geq p \geq 0}$\\

then \\

$\displaystyle{\mathbb{T}_{1}\mathbb{S}_{1,p} = \mathbb{I}_{\mathbb{E}_{p}^{\alpha}}}$ and $\displaystyle{\mathbb{T}_{2}\mathbb{S}_{2,p} = \mathbb{I}_{\mathbb{B}_{p}}}$ $\hfill { } (2.13)$\\

a) As $\displaystyle{\mathbb{T}_{1}^{k}e_{m,p}^{\alpha} = 0}$ for all $k > m \geq p$ and $\displaystyle{\mathbb{T}_{2}^{k}e_{n}^{p} = 0}$ for all $k > n \geq p$ we deduce that
any element of $\mathbb{F}_{\alpha}$ can be be annihilated by a finite power $k_{m}$ of $\mathbb{T}_{1}$ and any element of $\mathbb{F}$ can be be annihilated by a finite power $k_{n}$ of $\mathbb{T}_{2}$ \\

b) Since as $\displaystyle{[\prod_{j=m}^{k_{m}+m}\gamma_{j}^{\alpha,p}]^{-1}}$ and since as $\displaystyle{[\prod_{j=n}^{k_{n}+n}\omega_{j}^{p}]^{-1}}$ we get\\

$\left\{\begin{array}[c]{l}\displaystyle{\mathbb{S}_{1,p}e_{m,p}^{\alpha}=[\prod_{j=m}^{k_{m}+m}\gamma_{j}^{\alpha,p}]^{-1}e_{k + m,p}^{\alpha} \rightarrow 0 }$ in $\mathbb{E}_{p}^{\alpha}\\ \displaystyle{\mathbb{S}_{2,p}e_{n}^{p}=[\prod_{j=n}^{k_{n}+n}\omega_{j}^{p}]^{-1}e_{k + n}^{p} \rightarrow 0 }$ in $\mathbb{B}_{p}\\ \end{array}\right.\hfill { } (2.14)$\\

Now let $\displaystyle{\mathbb{S} = \mathbb{S}_{1,p}\otimes\mathbb{S}_{2,p}}$ and $\displaystyle{\mathbb{G} = \mathbb{F}_{\alpha}\otimes \mathbb{F}}$ which is dense in $\mathbb{E}_{p}^{\alpha}\otimes \mathbb{B}_{p}$ then \\

from (2.13) we deduce that\\

$\displaystyle{\mathbb{T}\mathbb{S} = \mathbb{I}_{\mid\mathbb{F}_{\alpha}\otimes \mathbb{F}}}$ $\hfill { } (2.15)\bigcup$\\

As the single tensor $0\otimes 0$ coincides with $\displaystyle{e_{m}^{\alpha}\otimes 0}$ and
$\displaystyle{0\otimes e_{n}^{p}}$ then for all $k > Min(m,n) \geq p $ in particular for $k > Max(m,n) \geq p $ we have $\displaystyle{\mathbb{T}^{k}e_{m,p}^{\alpha}\otimes e_{n}^{p} = 0}$.\\

then from (2.14), we deduce that any element of $\mathbb{G}$ can be annihilated by a finite power
$k_{m,n} = Max(k_{m}, k_{n})$ of $\mathbb{T}$ and \\

$\displaystyle{S^{k_{m,n}}e_{m,p}^{\alpha}\otimes e_{n}^{p} = [\prod_{j=m}^{k_{m,n}+m}\gamma_{j}^{\alpha,p}]^{-1}[\prod_{j=n}^{k_{m,n}+n}\omega_{j}^{p}]^{-1}e_{k+m,p}^{\alpha}\otimes e_{n}^{p}\rightarrow 0 }$ in $\mathbb{E}_{p}^{\alpha}\otimes \mathbb{B}_{p}$ $\hfill { } (2.16)$\\

Now, the hypercyclicity of $\mathbb{T}$ follows from the theorem of B$\grave{e}$s et al. recalled above.\\

\quad\\

{\bf{\color{blue} Lemma 2.9}}\\

{\it Let $\mathbb{T} = \mathbb{T}_{1}\otimes\mathbb{T}_{2}$ with domain $D(\mathbb{T}) = D(\mathbb{T}_{1})\otimes D(\mathbb{T}_{2})$ where\\

$\mathbb{T}_{1}= \mathbb{M}^{*^{p}}\mathbb{M}^{p+1}$ with domain $D(\mathbb{T}_{1}) = \{\psi \in \mathbb{E}_{p}^{\alpha}; \mathbb{T}_{1}\psi \in \mathbb{E}_{p}^{\alpha}\}$\\

$\mathbb{T}_{2}= z^{p}\frac{d^{p+1}}{dz^{p+1}}$ with domain $D(\mathbb{T}_{2}) = \{\phi \in \mathbb{B}_{p}; \mathbb{T}_{2}\phi \in \mathbb{B}_{p}\}$\\

Then\\

There exist $k > 0$ and $g \in D(\mathbb{T}^{k})$ such that $\mathbb{T}^{k}g = g$.}{\bf{\color{blue}{\hf}}}\\

{\bf Proof}\\

Let $(\lambda, \mu) \in \mathbb{C}^{2}$ and \\

$\displaystyle{g_{\lambda, \mu} = e_{p,p}^{\alpha}\otimes e_{p}^{p} + \sum_{m=p+1}^{\infty}\sum_{n=p+1}^{\infty}\frac{\lambda^{m-p}\mu^{n-p}}{(\gamma_{p}^{\alpha,p}...\gamma_{m-1}^{\alpha,p})(\omega_{p}^{p}...\omega_{n-1}^{p})}e_{m,p}^{\alpha}\otimes e_{n}}$ $\hfill { } (2.17)$\\

Then $\displaystyle{g_{\lambda, \mu} \in D(\mathbb{T})}$ and it is an eigenvector of $\mathbb{T}$ associated to eigenvalue $\lambda\mu$\\

In fact\\

Let $r > 0$ and $\mid \lambda\mu \mid < r$ then $\mid \lambda \mid < r$ and $\mid \mu \mid < r$\\

Now, as \\

$\displaystyle{Lim \prod_{j=p}^{m-1}\gamma_{j}^{\alpha,p} = +\infty}$ , $m \rightarrow +\infty $ $\hfill { } (2.18)$\\

and\\

$\displaystyle{Lim \prod_{j=p}^{n-1}\omega_{j}^{p} = +\infty}$, $ n \rightarrow +\infty$ $\hfill { } (2.19)$\\

then there exist $m_{0}, n_{0} \in N, q < 1$ and $q' < 1$ such that \\

$\displaystyle{\frac{r}{(\gamma_{p}^{\alpha,p}...\gamma_{m-1}^{\alpha,p})^{\frac{1}{m}}} \leq q}$  for $ m\geq m_{0}$ $\hfill { } (2.20)$\\

and \\

$\displaystyle{\frac{r}{(\omega_{p}^{p}...\omega_{n-1}^{p})^{\frac{1}{n}}} \leq q'}$ for $n\geq n_{0}$$\hfill { } (2.21)$\\

As  $\mid \lambda \mid < r$ and $\mid \mu \mid < r$, we deduce that\\

$\displaystyle{\frac{\mid \lambda \mid^{m-p}}{(\gamma_{p}^{\alpha,p}...\gamma_{m-1}^{\alpha,p})^{2}} \leq q^{2m}}$ and $\displaystyle{\frac{\mid \mu \mid^{n-p}}{(\omega_{p}^{p}...\omega_{n-1}^{p})^{2}} \leq q'^{2n}}$ for $ m\geq m_{0}, n\geq n_{0}$ respectively.\\

As $\{e_{m,p}^{\alpha}\otimes e_{n}^{p}\}$ is orthonoramal basis and \\

$\displaystyle{\sum_{m=p+1}^{\infty}\sum_{n=p+1}^{\infty}\frac{\mid\lambda\mid^{m-p}\mid\mu\mid^{n-p}}{(\gamma_{p}^{\alpha,p}...\gamma_{m-1}^{\alpha,p})^{2}(\omega_{p}^{p}...\omega_{n-1}^{p})^{2}} < \frac{(qq')^{p+1}}{(1-q^{2})(1-q'^{2})}}$\\

then $\displaystyle{g_{\lambda, \mu} \in E_{p}^{\alpha}\otimes B_{p}}$.\\

Now as \\

$\displaystyle{< g_{\lambda, \mu}, e_{p,p}^{\alpha}\otimes e_{p}^{p}>_{E_{p}^{\alpha}\otimes B_{p}} = 1}$\\

$\displaystyle{< g_{\lambda, \mu}, e_{k+1,p}^{\alpha}\otimes e_{k+1}^{p} >_{E_{p}^{\alpha}\otimes B_{p}} = \frac{\lambda^{m-p}\mu^{n-p}}{(\gamma_{p}^{\alpha,p}...\gamma_{m-1}^{\alpha,p})(\omega_{p}^{p}...\omega_{n-1}^{p})}}$ $\hfill { } (2.22)$\\

we get\\

$\displaystyle{\mid < g_{\lambda, \mu}, e_{k+1,p}^{\alpha}\otimes e_{k+1}^{p} >_{E_{p}^{\alpha}\otimes B_{p}}\mid^{2} = \frac{\lambda^{2(m-p)}\mu^{2(k-p)}}{(\gamma_{p}^{\alpha,p}...\gamma_{k}^{\alpha,p})^{2}(\omega_{p}^{p}...\omega_{
k}^{p})^{2}}}$ $\hfill { } (2.23)$\\

and\\

$\displaystyle{\mid < g_{\lambda, \mu}, e_{k+1,p}^{\alpha}\otimes e_{k+1}^{p} >_{E_{p}^{\alpha}\otimes B_{p}}\mid^{2} (\gamma_{k}^{\alpha,p})^{2}(\omega_{k}^{p})^{2} = \frac{\mid\lambda\mid^{2(k-p)}\mid\mu\mid^{2(k-p)}}{(\gamma_{p}^{\alpha,p}...\gamma_{k-1}^{\alpha,p})^{2}(\omega_{p}^{p}...\omega_{k-1}^{p})^{2}}}$ $\hfill { } (2.24)$\\

and \\

$\displaystyle{\mid < g_{\lambda, \mu}, e_{k+1,p}^{\alpha}\otimes e_{k+1}^{p} >_{E_{p}^{\alpha}\otimes B_{p}}\mid^{2} (\gamma_{k}^{\alpha,p})^{2}(\omega_{k}^{p})^{2} \leq (qq')^{2k}\mid\lambda\mu\mid^{2}}$ $\hfill { } (2.25)$\\

Then we deduce that $\displaystyle{ g_{\lambda, \mu} \in D(\mathbb{T})}$\\

Now as $\displaystyle{\mathbb{T} = \mathbb{T}_{1}\otimes \mathbb{T}_{2}}$ then \\

$\mathbb{T}g_{\lambda, \mu} = \mathbb{T}_{1}\otimes \mathbb{T}_{2}g_{\lambda, \mu} = \lambda\mu g_{\lambda, \mu}$ $\hfill { } (2.26)$\\

Therefore $g_{\lambda, \mu}$ is the eigenvector of $\mathbb{T}$ corresponding to the eigenvalue $\lambda \mu$ and it is a periodic point of $\mathbb{T}$ where $\lambda \mu$ is root of unity.{\bf{\color{blue}{\hf}}}\\

{\bf{\color{blue} Lemma 2.9}}\\

The set of periodic points of $\mathbb{T}$ is dense in $\displaystyle{\mathbb{E}_{p}^{\alpha}\otimes \mathbb{B}_{p}}$.\\

{\bf Proof}\\

Let $\displaystyle{\mathbb{U}_{per,\alpha} \subset \mathbb{E}_{p}^{\alpha}}$ and $\displaystyle{\mathbb{U}_{per,p} \subset \mathbb{B}_{p}}$ be the sets of periodic points for $\mathbb{T}_{1}$ and $\mathbb{T}_{2}$ respectively.\\

Let $\displaystyle{\mathbb{U}_{per} = \mathbb{U}_{per,\alpha}\otimes\mathbb{U}_{per,p}}$ which is a subset of set of periodic points of $\mathbb{T}$. As $\mathbb{T}_{1}$ and $\mathbb{T}_{2}$ are chaotic then $\displaystyle{\mathbb{U}_{per,\alpha}}$ and $\displaystyle{\mathbb{U}_{per,p}}$ are dense in $\mathbb{E}_{p}^{\alpha}$ and $\mathbb{B}_{p}$ respectively hence $\displaystyle{\mathbb{U}_{per}}$ is a dense subspace of $\displaystyle{\mathbb{E}_{p}^{\alpha}\otimes \mathbb{B}_{p}}$. In particular, The set of periodic points of $\mathbb{T}$ is dense in $\displaystyle{\mathbb{E}_{p}^{\alpha}\otimes \mathbb{B}_{p}}$.\\

Now, the chaoticity of $\mathbb{T}$ follows from the above lemmas.\\

We would like to finish this work with the following remark\\

{\bf{\color{blue} Remark 2.10}}\\

\n Let $z = (z_{1},.........,z_{j}, ....,z_{n}) \in \mathbb{C}^{n}; z_{j} = x_{j} + i y_{j} \in \mathbb{C}, 1\leq j \leq n$\\
$\displaystyle{\mathbb{B} = \{\phi : \mathbb{C}^{n} \rightarrow \mathbb{C}\quad entire ; \int_{\mathbb{C}^{n}}\mid \phi(z)\mid^{2}e^{-\mid z\mid^{2}}\prod_{j=1}^{n}dx_{j}\prod_{j=1}^{n}dy_{j} < +\infty\}}$\\
$\displaystyle{\mathbb{A}_{j}\phi = \frac{\partial}{\partial z_{j}}\phi}$ with domain $\displaystyle{D(\mathbb{A}_{j}) = \{\phi \in \mathbb{B};  \mathbb{A}_{j}\phi \in \mathbb{B}\}, 1\leq j \leq n}$\\
$\displaystyle{\mathbb{A}_{j}^{*}\phi =  z_{j}\phi}$ with domain $\displaystyle{D(\mathbb{A}_{j}^{*}) = \{\phi \in \mathbb{B};  \mathbb{A}_{j}^{*}\phi \in \mathbb{B}\}, 1\leq j \leq n}$\\
$\displaystyle{\mathbb{T} = \sum_{j=1}^{n}\mathbb{A}_{j}^{*}(\mathbb{A}_{j} + \mathbb{A}_{j}^{*})\mathbb{A}_{j}}$ with domain $\displaystyle{D(\mathbb{T}) = \{\phi \in \mathbb{B};  \mathbb{T}\phi \in \mathbb{B}\}}$\\
$\displaystyle{\mathbb{B}_{j} = \{\phi_{j} : \mathbb{C} \rightarrow \mathbb{C}\quad entire ; \int_{\mathbb{C}}\mid \phi_{j}(z_{j})\mid^{2}e^{-\mid z_{j}\mid^{2}}dx_{j}dy_{j} < +\infty\}}$\\

\n As $\displaystyle{\mathbb{B} = \otimes_{j=1}^{n}\mathbb{B}_{j}}$ we can to write $\mathbb{T}$ under the following form\\
\n $\displaystyle{\mathbb{T} = \oplus_{j=1}^{n}\mathbb{T}_{j}}$ where $\displaystyle{\mathbb{T}_{j} = I_{1}\otimes......\otimes\mathbb{A}_{j}\otimes....\otimes I_{n}}$\\

\n We observe that $\mathbb{T}$ is neither bounded nor self adjoint operator and as the direct sum of two hypercyclic operators is not in general a hypercyclic operator, indeed, Salas [26] showed that there exist hypercyclic operators $\mathbb{T}_{1}$ and $\mathbb{T}_{1}$ such that the direct sum $\mathbb{T}_{1} \oplus \mathbb{T}_{2}$ is not hypercyclic.\\

\n We will give in another paper a comparison of chaoticity of direct sums with chaoticity of tensor products for these operators acting classical Bargmann space or for the operators acting on generalized Fock-Bargmann space like those definite in [16].\\

\begin{center}

{\bf{\Large {\color{blue}References}}}

\end{center}

\n [1] S. I. Ansari, Hypercyclic and cyclic vectors, Journal of Functional Analysis, vol. 128, no. 2, pp. 374–383\\

\n [2] J. Banks, J. Brooks, G. Cairns, G. Davis and P. Stacey, On Devaney's definition of chaos, The American Mathematical Monthly, vol. 99, no. 4, pp. 332–334, 1992.\\

\n [3] V. Bargmann, : "On a Hilbert space of analytic functions and an associated integral transform," Communications on Pure and Applied Mathematics, vol. 14, pp. 187–214, 1961.\\

\n [4] J. B$\grave{e}$s, K. Chan, and S. Seubert, Chaotic unbounded differentiation operators, Integral Equations Operators Theory, vol. 40, no. 3, pp. 257–267, 2001.\\

\n [5] M. Birkhoff, D\'emonstration d'un th\'eor$\grave{e}$me \'el\'ementaire sur les fonctions enti$\grave{e}$res, Comptes Rendus de l'Acad\'emie des Sciences, vol. 189, pp. 473–475, 1929.\\

\n [6] A. Decarreau, H. Emamirad and A. Intissar, "haoticit\'e de l'op\'erateur de Gribov dans l'espace de Bargmann, Comptes Rendus de l'Acad\'emie des Sciences, vol. 331, no. 9, pp. 751–756, 2000.\\

\n [7] R.L. Devaney, An Introduction to Chaotic Dynamical Systems, 2nd Edition, Addison-Wesley,
Reading, MA, 1989.\\

\n [8] G.B. Folland, A Course in abstract harmonic analysis, CRC Press BOCA Raton, Florida (1995)\\

\n [9] A. Ghanmi and A. Intissar, Construction of concrete orthonormal basis for $(L^{2}; \Gamma;  \chi)$-theta functions associated to discrete subgroups of rank one in $(\mathbb{C}; +)$ J. Math. Phys. 54, 063514 (2013)\\

\n[10] G. Godefroy, J.H. Shapiro, Operators with dense, invariant, cyclic vector manifolds, J. Funct. Anal. 98 (2) (1991) 229–269.\\

\n [11] K.G. Grosse-Erdmann, Universal families and hypercyclic operators, Bull. Amer.
Math. Soc. 36, (1999), 345-381\\

\n [12] K.G. Grosse-Erdmann, Hypercyclic and chaotic weighted shifts, Studia Math. 139,
(2000),4768\\

 \n [13] A. Gulisashvili, and C.R. MacCluer, "Linear chaos in the unforced quantum harmonic oscillator," Journal of Dynamic Systems, Measurement and Control, vol. 118, no. 2, pp. 337–338, 1996.\\

\n [14] A. Intissar, A short note on the chaoticity of a weight shift on concrete orthonormal basis associated to some Fock-Bargmann space, Journal of Mathematical Physics 55, 011502 (2014); doi: 10.1063/1.4861931\\

\n [15] A. Intissar, On a chaotic weighted shift $z^{p}\frac{d^{p}}{dz^{p}}$ of order $p$ in Bargmann space, Advances in Mathematical Physics,(2011)\\

\n [16] A. Intissar,  On a chaotic weighted shift in generalized Fock-Bargmann spaces, Math.
Aeterna, Vol.3, no.7, (2013) 519-534

\n [17] Intissar, "Analyse de scattering d'un op\'erateur cubique de Heun dans l'espace de Bargmann," Communications in Mathematical Physics, vol. 199, no. 2, (1998). 243–256.\\

\n A. [18] Intissar, Etude spectrale d'une famille d'op\'erateurs non-sym\'etriques intervenant dans la th\'eorie des champs de Reggeons, Comm. Math. Phys. 113 (1987) 263-297.\\

\n [19] A. Intissar, Spectral Analysis of Non-self-adjoint Jacobi-Gribov Operator and Asymptotic Analysis of Its Generalized Eigenvectors, Advances in Mathematics (China), V. 43, (2014) doi: 10.11845/sxjz.2013117b\\

\n [20] C.S. Kubrisly, A concise introduction to tensor product, Far East Journal of Mathematical Sciences 22 (2006) 137- 174 \\

\n [21] G.R. Maclane, Sequences of derivatives and normal families, Journal d'Analyse Math\'ematique, vol. 2, pp. 72–87, 1952.\\

\n [22] F. Martinez-Gim\'enez, and  A. Peris, Universality and chaos for tensor products of operators, Journal of Approximation Theory 124 (2003) 724\\

\n [23] Reed-Simon, tensor products of closed operators on Banach spaces, Journal of Functional Analysis, 13, (1973), 107-124]\\

 \n [24] H.N. Salas, Pathological hypercylic operators, Archiv der Mathematik, vol. 86, no. 3, pp. 241–250, 2006.\\

 \n [25] H.N. Salas, Hypercyclic weighted shifts, Trans. Amer. Math. Soc. 347 (3) (1995) 993–1004.\\

 \n [26] H.N. Salas, A hypercyclic operator whose adjoint is also hypercyclic, Proc. Amer. Math.
Soc. 112 (1991), 765-770.\\

\n [27] R. Schatten, A theory of cross-spaces, Annals of Mathematics studies, Number 26, Princeton University Press (1950)\\

\n [28] J. Weidmann, Linear Operators in Hilbert Spaces, Springer, New York, 1980.\\

\end{document}